\documentclass{article}
\usepackage[ansinew]{inputenc}

\usepackage[dvips]{graphicx}

\usepackage{url}
\usepackage{amsmath}

\usepackage{color}
 \usepackage{setspace}

 \usepackage[square,numbers, super, sort]{natbib}

\title{Bayesian classification for dating archaeological sites via projectile points}
\author{Carmen Armero.\\
 Departament d'Estad\'istica i IO. Universitat de Val\`encia. \and
 Gonzalo Garc\'ia-Donato.\\
  Department of Economics and Finance. Universidad Castilla-La Mancha. \and
 Joaqu\'in Jim\'enez-Puerto,  Salvador Pardo-Gord\'o, and Joan Bernabeu.\\ Department of Prehistory, Archaeology and Ancient History, Universitat de Val\`encia.}

\date{ }

\begin{document}

\maketitle

\begin{abstract}
   Dating is a key element for archaeologists. We propose a Bayesian
approach to provide chronology to sites that have neither radiocarbon dating nor
clear stratigraphy and whose only information comes from lithic arrowheads. This classifier is based on the Dirichlet-multinomial inferential process and
posterior predictive distributions. The procedure is applied to predict the period
of a set of undated sites located in the east of the Iberian Peninsula during the
IVth and IIIrd millennium cal. BC.\vspace*{-0.3cm}\\

\end{abstract}

\section{Introduction}\label{sec:introduction}

 Dating is a key element for archaeologists. A time scale to locate the information collected from  excavations and field work is always necessary  in order to build, albeit with uncertainty, our most remote past.
Archaeological scientists generally use stratigraphic expert information and dating techniques
for examining the age of the relevant artifacts. Bayesian inference is commonly used in archaeology as a tool to construct robust chronological models based on information from scientific data as well as expert knowledge (e.g. stratigraphy) (Buck el at., 1996).

Radiocarbon dating is one of the most popular   techniques for obtaining data due
to its presence in any being that has lived on Earth.  However, it is not always possible in all studies to collect organic material and obtain that type of data or to have good stratigraphic references.
In this context, we propose  a Bayesian approach   to provide chronology to some archaeological sites that do not have radiocarbon dates and show   unprecise stratigraphic relationships.

We propose an automatic Bayesian procedure, very popular in text classification (Wang \textit{et al.}, 2003), based on   predictive probability distributions for classifying the period to which an undated site belongs  according to the type and number of arrows found in it. This  proposal takes into account  the Dirichlet-multinomial inferential process for learning about the proportion of different types of arrowheads in each chronological period and the concept of posterior predictive distribution for a new undated site. This procedure is applied to  date a set of sites located in the east of the Iberian Peninsula during the IVth and IIIrd millennium cal. BC. During this time, bifacial flint arrowheads appear and spread. Archaeological research suggests that the shape of these arrowheads could be related with specific period and/or geographical social units spatially defined.

This paper is organized in five Sections.  Following this
introduction, Section 2 briefly introduces the archaeological framework and the   lithic material that will be the basis for the classification process. Section 3 describes the two stages of the Bayesian statistical analysis.   The first is of an inferential type and focuses on the study of the abundance of different types of arrows in the different periods considered.  The second uses the information from the first stage to predict the period of an undated deposit from the number and type of arrowheads encountered. Section 4 applies the methodological procedure from previous Section to a set of sites in the east of the Iberian Peninsula during Late Neolithic and Chalcolithic (IV-III millennium BC).   Finally, Section
5 concludes.

\section{Chronological periods and lithic information}

The arrival of the neolithic economy, based on domestic resources, is dated on the first half of the VI millennium cal BC. We will have to wait until the IV-III millennium to be able to witness clear winds of change. This is the moment of the appearance of a higher level of hierarchy in some societies. The Late Neolithic (IV-III millennium cal. BC) in the oriental Iberian fa\c{c}ade, is the time of the transit to a higher complexity in social and economic terms. This process will last long and it will crystallize by the end of the IIIrd millenium cal BC. (Bernabeu and Orozco, 2014). The evaluation of this process in such a huge frame faces some problems which need to be addressed. One of these difficulties is closely associated to the chronological attribution of a big part of the period's archaeological record.

One of the main  goals in archaeological research is focused on the way the members of the prehistoric cultures interact with the landscape and the objects. From an evolutive perspective, the way human cultures change through space-time is determined by inheritance patterns, adaptation and interaction (Shennan, Crema, and Kerig, 2015). Therefore, the analysis of items from the archaeological record, able to capture the cultural evolution of the human groups, would be a main goal for the researcher. Being able to assign non radiocarbon dated collections to a specific chronological lapse should be a very useful weapon in the archaeologist arsenal.

The concept of ``culture'' covers many factors. Hence, we will use the material culture as an archaeologic proxy in order to analyze the evolution and dispersion of the cultural traits in the study area. However, not all the items included in material culture are useful for that. Those which show a wide geographic and cultural dispersion or whose variability is low are not convenient to detect changes. So, we will use elements with markers pointing to the different cultural traits of the social groups involved,  as well as the social relationships between them. One of these useful items are the lithic productions, and more specifically the arrow heads.

 The classification of the arrowheads is based on the previous works performed around the typological formalization for the study area. They are mainly inspired by morfo-descriptive typologies. Therefore, the classification contains a functional and morphological meaning. Arrowheads constitute a very representative tool group of the Late Neolithic and Chalcolithic. Their function is quite proved thanks to the studies in traceology, experimental archaeology and etnoarchaeology. Some well known examples are the spectacular findings of arrow heads still nailed into the victim bones, present in many burials from the IV and III millennium BC (i.e. San Juan ante Portam Latinam: Vegas 2007). We cannot forget the awesome finding of a full equipment \"{O}tzi, the ``Iceman'', was carrying in the Alps (Cave-Browne, 2016), and exceptionally conserved. Moreover, the existence of excavated sites (Ereta del Pedregal) in which the whole arrowhead operative chain process can be observed, has provided additional information (Juan-Cabanilles, 1994).

 \begin{figure}[h]
\begin{center}
\includegraphics[width=16cm ]{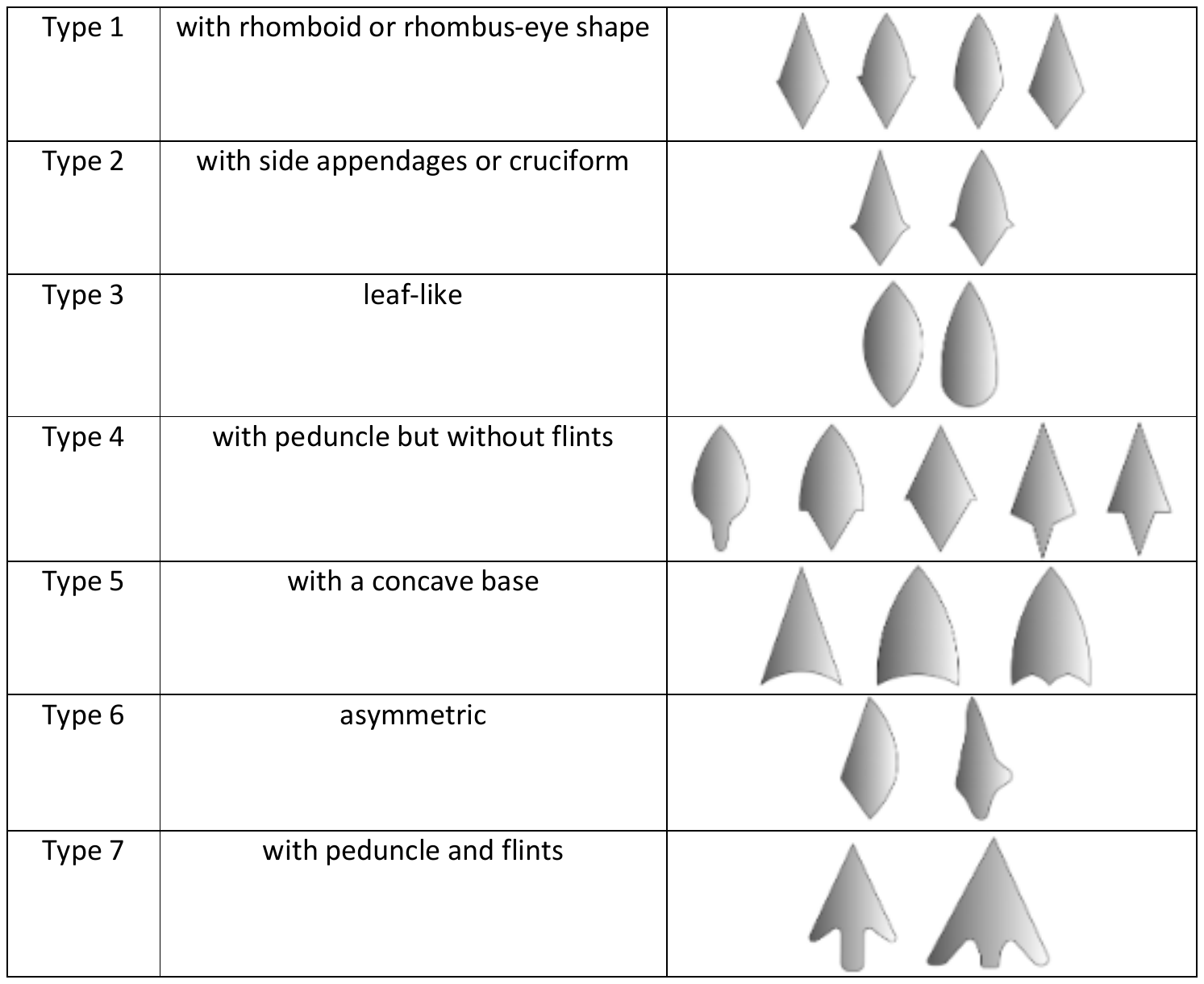}\vspace*{-10cm}\\
\caption{\label{fig:arrows} Arrow head types used for the study.}
\end{center}
\end{figure}

The arrowhead types present in the archaeological records have been classified in seven types  following a morphological criterion, based on previous typologies for the study area
(Juan-Cabanilles, 2008) (See    Figure \ref{fig:arrows}).

 \section{Bayesian classification process}

Bayesian classification within the framework of archaelogical datation with lithic information will provide  a probability distribution for the period   to which an undated site  belongs    in which  a given set of different types of arrowheads has been found. This probability distribution depends on the knowledge of the abundance  of each type of arrowheads in each period, expressed via the posterior distribution for the probability associated with each type of arrowhead,  and the posterior predictive distribution for the period of that particular updated site.

\subsection{Dirichlet-multinomial inferential process}

 Let $Y_{ij}$ be the random variable that   describes the number of type $j$, $j=1,\ldots, J$ arrowheads, of the total $n_i$ collected  in the sites belonging to period $i$, $i=1, \ldots,I$. We define the random vector
 $\boldsymbol{Y}_{i}=(Y_{i1}, Y_{i2}, \ldots, Y_{i,J-1})^{\prime}$ and the probability vector $\boldsymbol \theta_i= (\theta_{i1}, \theta_{i2}, \ldots, \theta_{i,J-1})^{\prime}$, where $\theta_{ij}$ is the probability  that an arrowhead  of period $i$ is of type $j$. A probabilistic model for $\boldsymbol{Y}_{i} \mid \boldsymbol \theta_i$  is the multinomial distribution,  $\mbox{Mn}(\boldsymbol \theta_i, n_i)$, with probability distribution

 \begin{equation}
 f(\boldsymbol y_{i}  \mid \boldsymbol \theta_i)= \frac{n_i !}{\Big (\prod_{j=1}^{J-1}\,y_{ij}! \Big )\,y_{iJ}!}\,\Big(\prod_{j=1}^{J-1}\,\theta_{ij}^{y_{ij}} \Big )\,\theta_{iJ}^{y_{ij}},
 \end{equation}

 \noindent where $\boldsymbol y_{i}$ is an observation of $\boldsymbol Y_{i}$, $y_{iJ}=n_i- \sum_{j=1}^{J-1}\, y_{ij}$ is the total number of arrowheads of type $J$ in the sites of period $i$ and
  $\theta_{iJ}=1-\sum_{j=1}^{J-1}\, \theta_{ij}$ is the probability  that an arrowhead of period $i$ is of type $J$.

The combination of a multinomial sampling  model with a conjugate Dirichlet prior distribution was
proposed by Lindley (1964) and Good (1967) as the generalisation of the beta-binomial model. The Dirichlet distribution for $\boldsymbol \theta_i$ with parameters $\boldsymbol \alpha_i =(\alpha_{i1}, \ldots, \alpha_{iJ})^{\prime}$, $\alpha_{ij}>0, j=1, \ldots, J$, Dir$(\boldsymbol \alpha_i)$,  is a multivariate continuous distribution with joint density function
\begin{equation}
\pi(\boldsymbol \theta_{i})= \frac{\Gamma(\alpha_{i+})}{\prod_{j=1}^{J}\,\Gamma(\alpha_{im})} \, \Big ( \prod_{j=1}^{J-1}\, \theta_{ij}^{\alpha_{ij}-1} \Big ) \, \theta_{iJ}^{\alpha_{iJ}-1},
\end{equation}

\noindent where $\Gamma(\cdot)$ represents the gamma function and $\alpha_{i+}=\sum_{j=1}^{J}\,\alpha_{ij}$.

\begin{figure}[h]
\begin{center}
\includegraphics[width=14cm ]{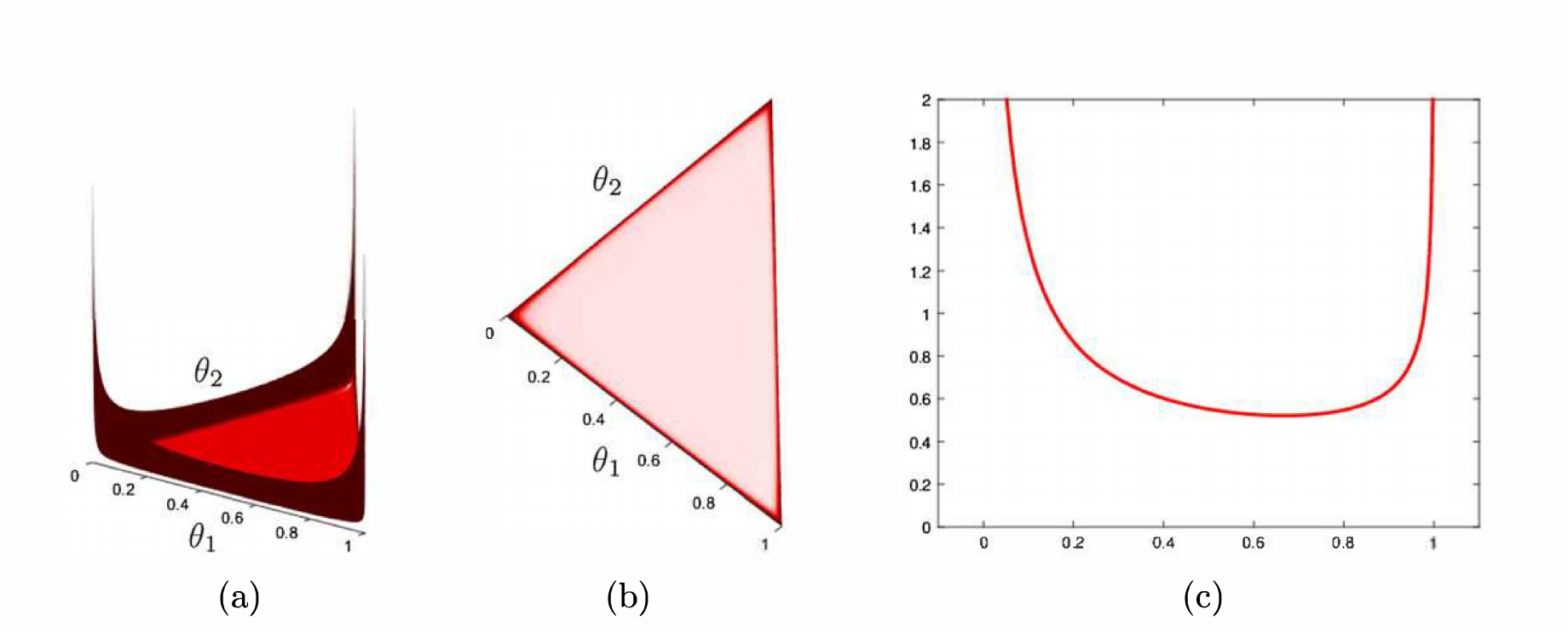}
\caption{Perks' distribution  when the number of  types of arrowheads is $J=3$  (a), its projection onto the simplex triangle (b), and the  marginal prior distribution
for each individual component, a beta distribution with parameters $1/3$ and $2/3$, Be($1/3$, $2/3$), which maintains high density values close to 0 and 1(c) (Figure from Alvares et al., 2018).\label{fig:Perk} }
\end{center}
\end{figure}

We assume an inferential process for each  $\boldsymbol \theta_i,\,i=1,\ldots,I$ in the framework of   the  Dirichlet-multinomial  process  with a non-informative prior  distribution for $\boldsymbol \theta_i$ that gives all the protagonism of the  process to the data. There are many proposals for elicit the parameters $\boldsymbol \alpha_i$ in a non-informative way: Haldane's prior, Perks' prior or reference distance prior, hierarchical approach prior and Jeffreys' prior or common reference prior, and Bayes-Laplace prior. All them have good theoretical properties but they also have some small shortcomings. We choose the Perks' prior as a result of   Alvares et al (2018). This prior was firstly proposed by Perks (1947),
but recently it has been also obtained as the reference distance prior by Berger et al. (2015).
This is a Dirichlet distribution with all  parameters  equal to $1/J$,  where  $J$ is the number of arrow types.
Figure  \ref{fig:Perk} shows the density  and other characteristics of a Perk's distribution with three categories.

The posterior distribution for $\boldsymbol \theta_i$ when data $\boldsymbol y_{i}$ are
observed is also a Dirichlet distribution,
\begin{equation}
\pi(\boldsymbol \theta_{i} \mid \boldsymbol y_i)= \mbox{Dir}(\alpha_{i1}=y_{i1}+(1/J),  \ldots, \alpha_{iJ}=y_{iJ}+(1/J)).
\end{equation}

This posterior distribution has an important feature: never assigns absolute probabilities 1 or 0 to the presence of  any type   of headarrows.
  The marginal posterior distribution for each probability $\theta_{ij}$ is the beta distribution
  \begin{equation}
  \pi(\theta_{ij} \mid \boldsymbol y_i)=\mbox{Be}(\alpha_{ij}, \alpha_{i+}-\alpha_{ij}),
 \end{equation}
with posterior mean and variance $\alpha_{ij}/ \alpha_{i+} $ and  $\alpha_{ij} (\alpha_{i+}-\alpha_{ij})/(\alpha_{i+}^2 \, (\alpha_{i+}+1))$, respectively.

  \subsection{Predictive process}

 After learning about the distribution of the proportion of arrowheads types in each site, we   have to assign a probability distribution to the period $m^*$  to which a new undated
 site $s^*$ belongs given that  a total of $n^*$ arrowheads $\boldsymbol y^{*}=(y^{*}_1,  \ldots, y^{*}_J)$ have been found in it. Following Bayes' theorem:
\begin{equation}
P(m^*=m_i \mid \boldsymbol y^{*}, \boldsymbol y) \propto P(\boldsymbol Y^{*}=\boldsymbol y^{*} \mid m^*=m_i, \boldsymbol y)\,
P(m^*=m_i   \mid \boldsymbol y), \,\,i=1,\ldots, I
\label{eqn:predictive}
\end{equation}
where $ \boldsymbol y=(\boldsymbol y_1, \ldots, \boldsymbol y_I)^{\prime}$ and $\boldsymbol Y^*=(Y^*_1, \ldots, Y^*_J)$ is the random vector that describes the number of arrowheads of the different types that will be recorded in that new site.

The posterior predictive distribution in (\ref{eqn:predictive}) is proportional to the product of two terms. The first one is:

\begin{align*}
P(\boldsymbol Y^{*}= \boldsymbol y^{*} \mid & m^*=m_i, \boldsymbol y ) =   \int P(\boldsymbol Y^{*}=\boldsymbol y^{*} \mid  \boldsymbol \theta_i, m^*=m_i, \boldsymbol y)\, \pi(\boldsymbol \theta_{i} \mid m^*=m_i,\boldsymbol y) \,\mbox{d}\boldsymbol \theta_i \\
    & = \int  \frac{n^*!}{y_1^*! \, y_2^{*}! \cdots y_J^*!} \, \theta_{i1}^{y_1^*} \,\theta_{i2}^{y_2^*} \cdots  \theta_{iJ}^{y_J^*}  \,\frac{ \Gamma(\alpha_{i+})}{ \prod_{j=1}^{J}\,\Gamma(\alpha_{ij})} \,\theta_{i1}^{\alpha_{i1}-1} \,\theta_{i2}^{\alpha_{i2}-1} \cdots  \theta_{iJ}^{\alpha_{iJ}-1} \mbox{d} \boldsymbol \theta_i \\
    & = \frac{n^*!}{y_1^*! \, y_2^{*}! \cdots y_J^*!} \, \frac{ \Gamma(\alpha_{i+})}{ \prod_{j=1}^{J}\,\Gamma(\alpha_{ij})} \int \,   \,\theta_{i1}^{\alpha_{i1}+y_1^*-1} \,\theta_{i2}^{\alpha_{i2}+y_2^*-1} \cdots  \theta_{iJ}^{\alpha_{iJ}+y_J^*-1} \mbox{d} \boldsymbol \theta_i\\
      & = \frac{n^*!}{y_1^*! \, y_2^{*}! \cdots y_J^*!} \, \frac{ \Gamma(\alpha_{i+})}{ \Gamma(\alpha_{i+}+n^*)}
       \prod_{j=1}^{J}\,\frac{\Gamma(\alpha_{ij}+y_j^*)}{\Gamma(\alpha_{ij}).}
\end{align*}

The second element in the product in  (\ref{eqn:predictive}),   $P(m^*=m_i \mid \boldsymbol y)$, can be estimated as the proportion of sites in the sample for each of the periods under consideration (Barber, 2012).

\section{East of the Iberian Peninsula sites during the IVth and IIIrd millennium cal. BC.}

We apply the classification procedure above to a set of undated sites in the East of the Iberian Peninsula during the the IVth and IIIrd millennium cal. BC. Data for the inferential process of the study come from  31 archaeological sites radiocarbon dated with arrowheads, clear contexts and stratigraphy.

\subsection{Inferential process}

All 14C dated sites have been filtered   using only  those whose radiocarbon dates come from short-lived singular samples.   The final levels used for the periodization are:
Arenal de la Costa (Bernabeu, 1993), Barranc del Migdia (Soler \textit{et al.}, 2016), Beniteixir (Pascual-Beneyto, 2010), Cam\'i de Missena (Pascual-Beneyto \textit{et al.}, 2005), Colata (G\'omez-Puche \textit{et al.}, 2004), Cova del Randero (Soler \textit{et al.}, 2016), Cova dels Diablets (Aguilella \textit{et al.}, 1999), Jovades (Bernabeu, 1993), La Vital (P\'erez-Jord\'a \textit{et al.}, 2011), Niuet (Bernabeu \textit{et al.}, 1994), and Quintaret (Garc\'ia-Puchol \textit{et al.}, 2014). These sites are
 located in the eastern Mediterranean area. Figure \ref{fig:iberia} shows a map with the dated sited as well as the sites without 14C  datation whose time classification is the final object of this study.

\begin{figure}[h]
\begin{center}
\includegraphics[width=12cm ]{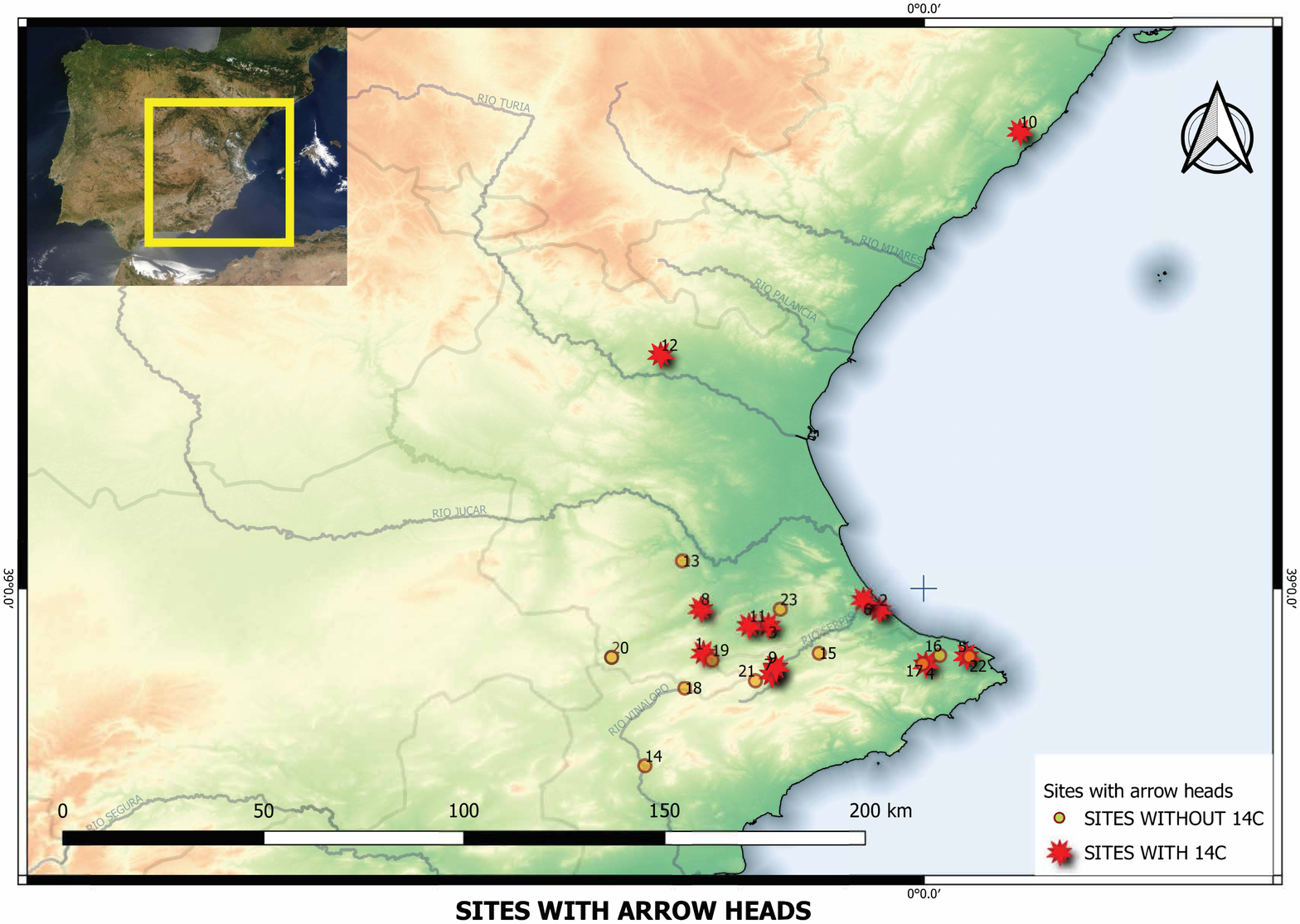}
\caption{Situation map of the sites with arrow heads present in the study area.\label{fig:iberia} }
\end{center}
\end{figure}
Based on the chrono-stratigraphic  and available expert information, we have proposed   five intervals or chronological periods organization comprised between ca. 4600-3200 cal BC.
Table \ref{tab:14Cdated}  includes the period and duration of each of the periods considered as well as   the sites included in each of them.

Each site usually contains many different archaeological levels attached to different moments of occupation. In this specific case, archaeological contexts containing arrowheads have been dated through
radiocarbon determinations. Some of these sites contain different dated levels in which arrowheads were present. Hence we have described them with the name of the site and a number to differentiate them.
Based on the chrono-stratigraphic and available expert information, we have proposed five intervals or chronological periods organization comprised between ca. 4600-2150 cal BC. These periods have
 resulted from the application of Bayesian radiocarbon modeling methods to the archaeologic  information available for each period.

 Table 1 includes the period of each of the periods considered as well as the sites included in each of them.

\begin{table}[h!]
\begin{center}
\caption{Periods and sites extracted from clear archaeological contexts with radiocarbon determinations. }\label{tab:14Cdated}
 \begin{tabular}{lc}
Sites  14C dated & Period \\
\hline
Jovades 1, Jovades 2, and Niuet 1 &   1	 \\
Colata 1, Colata 2, Jovades 3, Jovades 4, Niuet 2,   &   2  \\
and Quintaret &    	\vspace*{0.2cm}\\
Beniteixir, Diablets 1, Diablets 2,  Diablets 3,     &   3 \\
Jovades 5, La Vital 1,  La Vital 2, Migdia 1,        &     \\
Missena 1,   Niuet 3, Niuet 4, Randero 1,            &     \\
and Randero 2                                        &     \\
La Vital 3, Migdia 2, Missena 2,   and Missena 3     &  4  \\
Arenal Costa, La Vital 3, 	Missena 4, Missena 5,    &  5   	  \\
and Missena 6                                        &       \\
\hline
\end{tabular}
\end{center}
\end{table}

Table \ref{tab:jointposterior} includes the posterior distribution of the different types of arrowheads in each of the periods considered.  Although we have already commented on this, we would like to emphasise once again that the 1/7 values associated with the parameters of each Dirichlet posterior correspond to types of arrowheads not present in the subsequent sample.  They are small values that avoid absolute probabilities of zero, impossible to update in the case of an additional incorporation of information to the inferential process.

\begin{table}[h!]
\begin{center}
\caption{Posterior Dirichlet distribution for the proportion of arrowheads from type 1 to type 7 in each of the periods considered. }\label{tab:jointposterior}
 \begin{tabular}{cl}
Period& Posterior distribution \\
\hline
  1	& Dir(15/7, 22/7, 8/7, 1/7, 1/7, 1/7, 1/7) \\
  2	& Dir(29/7, 36/7, 15/7, 8/7, 1/7, 1/7, 1/7) \\
  3	& Dir(43/7, 1/7, 43/7, 64/7, 29/7, 1/7, 71/7) \\
  4	& Dir(15/7, 1/7, 15/7, 8/7, 15/7, 1/7, 43/7) \\
  5	& Dir(1/7, 1/7, 1/7, 15/7, 1/7, 8/7, 36/7) \\

\hline
\end{tabular}
\end{center}
\end{table}

\begin{figure}[h]
\begin{center}
\includegraphics[width=14cm]{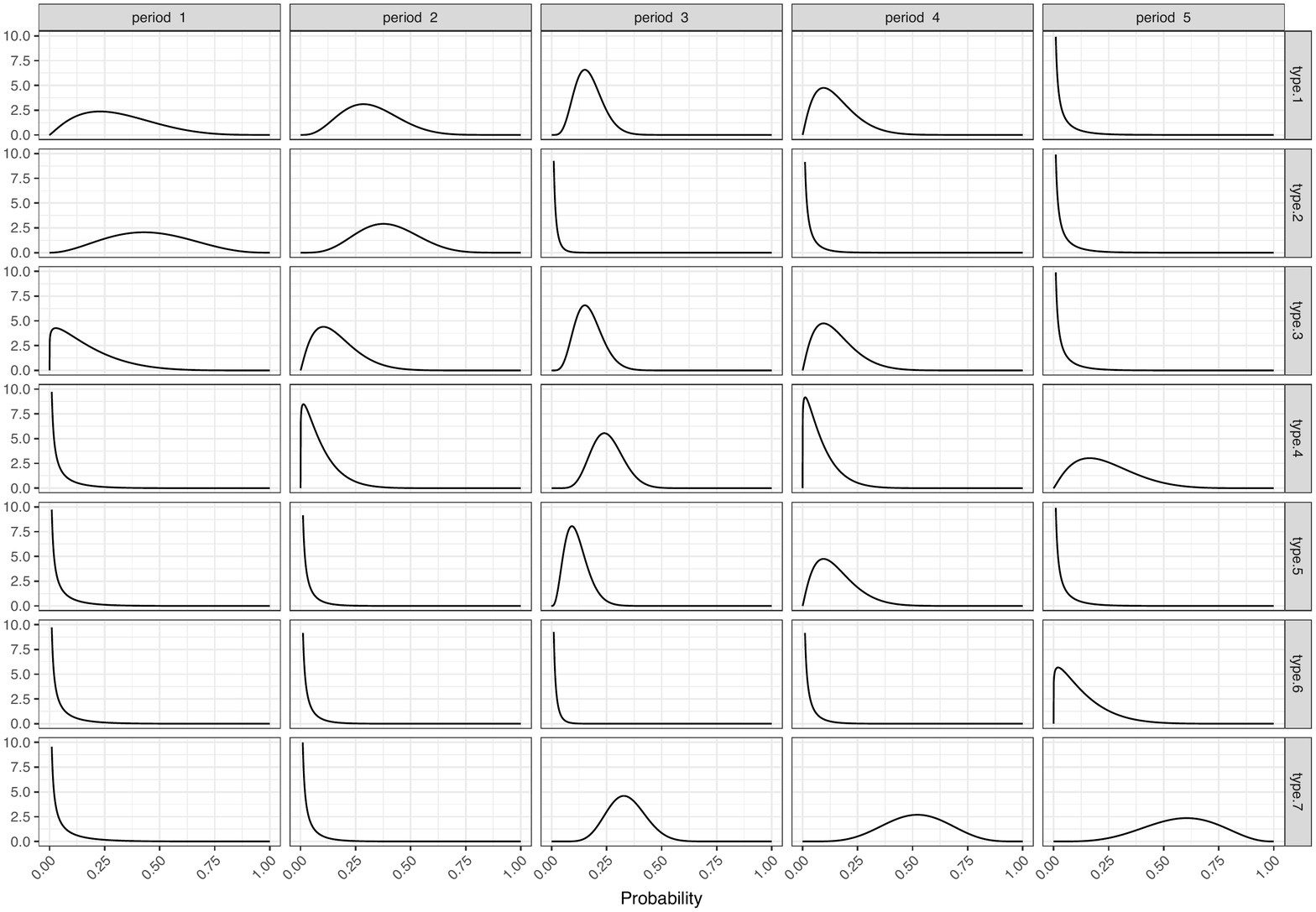}\\
 \caption{ Posterior marginal distribution  for the probability associated with each type of arrowhead in each of the periods in the study. } \label{fig:periodarrow}
\end{center}
\end{figure}

Table \ref{tab:posteriormean} shows the posterior mean  for the probability associated with each type of arrowhead in each of the periods in the study.
Figure \ref{fig:periodarrow} shows the posterior marginal distribution of the abundance of the   different types of arrowheads in each of the five chronological periods considered. Results in Table \ref{tab:posteriormean} and Figure \ref{fig:periodarrow} indicate that the distribution of the different types of arrowheads is very similar in    Periods 1 and 2:  Type 1 and 2 arrowheads are the most abundant and   about the 75$\%$ and $\%70\%$ of the total of arrowheads in both periods are type 1 or 2.  Type 3 arrowheads   have  poor relevance in both Periods and types 4, 5, 6, and 7 are virtually nonexistent. In Period 3 we find practically no type 2 and 6  arrowheads.  The rest of arrowheads in this period  have a presence quite similar but type 4 and 7 have a slightly higher presence. Period 4 shows a large presence of type 7 arrows and, to a lesser extent, of type 1, 3 and 5 arrows (probabilities of about 0.15). Arrowheads of type 2 and 6 have no relevance. Approximately 57$\%$ and 24$\%$   of the arrows of Period 5 are of type 7 and 4, respectively. The rest of the arrowheads types, except possibly those of type 6, are essentially irrelevant.

\begin{table}[h!]
\begin{center}
\caption{Posterior mean of the probability  of abundance of each type of arrowhead in each of the periods of the study}{\label{tab:posteriormean}}
 \begin{tabular}{clllll}
Type & Period 1 & Period 2& Period 3 & Period 4 & Period 5\\
\hline
  1    & 0.3061 & 0.3187 & 0.1706 & 0.1531 & 0.0159\\
  2    & 0.4490 & 0.3956 & 0.0040 & 0.0102 & 0.0159\\
  3    & 0.1633 & 0.1648 & 0.1706 & 0.1531 & 0.0159\\
  4    & 0.0204 & 0.0879 & 0.2540 & 0.0816 & 0.2380\\
  5    & 0.0204 & 0.0110 & 0.1151 & 0.1531 & 0.0159\\
  6    & 0.0204 & 0.0110 & 0.0040 & 0.0102 & 0.1270\\
  7    & 0.0204 & 0.0110 & 0.2817 & 0.4387 & 0.5714\\
\hline
\end{tabular}
\end{center}
\end{table}

\subsection{Predictive process}

Undated sites  between  the IVth and IIIrd millennium cal. BC. used to explore the predictive approach include burial sites, villages, and
caves: Barranc Cafer 2, Barranc Parra 3,
Casa Color\`a, Cova Ampla del Montg'o, Cova Santa Vallada B, Cova de les Aranyes, Cova dels Anells,
Cova del Negre, Cova del Petrol\'i, Cova Pardo, Cova Santa Vallada A, Ereta I, Ereta II, Ereta III,
Ereta IV, Escurrupenia, Font de Mahiques, Garrofer 3, Garrofer K, Garrofer I-J, Rambla C.,
Sima de la Pedrera, Niuet s3, Torreta UE1, and Torreta UE2 (See Figure \label{fig:iberia}).

The posterior probability that a new site   belongs to each of the periods considered was estimated as 0.15 for Periods 1, 4 and 5, 0.20 for Period 2, and 0.35 for Period 3.

\begin{figure}[h]
\begin{center}
\includegraphics[width=14cm]{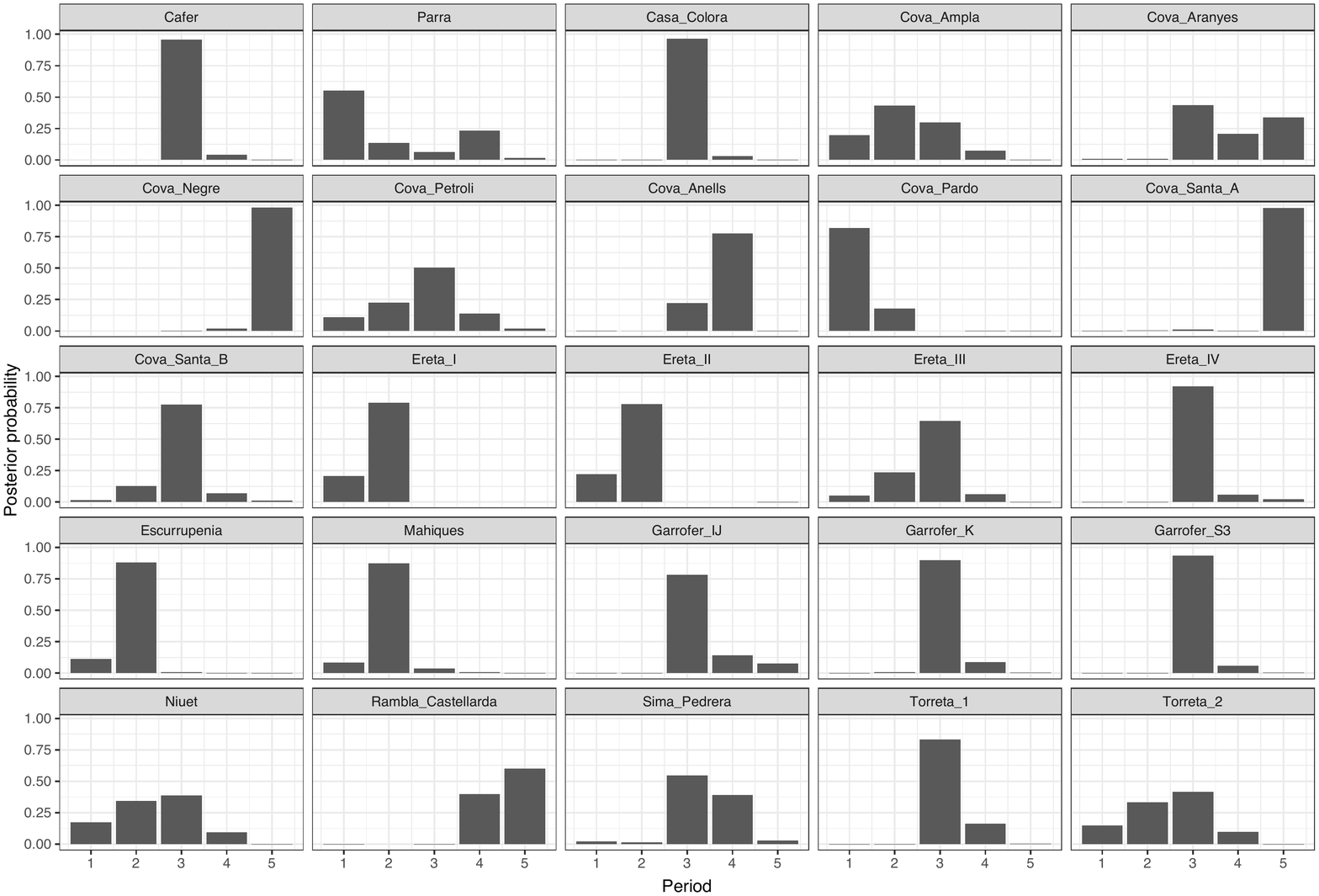}\\
 \caption{ Posterior marginal distribution  for the probability associated with each type of arrowhead in each of the periods in the study. } \label{fig:periodsarrows}
\end{center}
\end{figure}

Figure \ref{fig:periodsarrows} presents the   posterior  predictive distribution of the period to which the above  undated sites belong, whose only available information is
 based on the number and type of arrows found collected.

The results obtained show a high concordance with the expert information provided by archaeologists. Thus, for example, in those sites that present stratigraphic correlations (Ereta del Pedregal and La Torreta) the chronological evaluation obtained from the predictive approach is consistent with the chrono-statigraphical information. The case of Cova Santa de Vallada B is interesting, which from the archaeological information is situated in phase 3-4. However, based on Bayesian modeling, this indicates that it should be located in phase 3. This aspect is totally coherent not only because of the typology of the arrowheads themselves but also because of the presence of other diagnostic elements such as the presence of metal and the absence of bell-beaker ceramics. The result is totally consistent with the cases of Casa Color\`a and Cova del Garrofer I-J, which both the previous experience and the Bayesian application place in phase 3. Finally, there are some cases in which the results qualify the chronological proposal established by expert knowledge, such as the case of Barranc de Parra 3, where previous knowledge places it in phase 2-3 but predictive analysis places it either in phase 1 or in phase 4. In this sense, we must bear in mind both that there may be a persistence of certain types of arrowheads throughout the entire sequence analyzed, as is the case of the arrowheads of the peduncle, as well as the possible reuse of projectiles located in places of habitat as has been documented in the Clovis culture, North America. In this sense both the incorporation of other complementary diagnostic archaeological information (presence of metal and bell-shaped ceramics) may help to establish a more precise chronology.

 \section*{Conclussions}

In short, results obtained present a good agreement with the expert information of the archaeologists, so it is a proposal that can be very useful in archaeological research.  However, there is no doubt that both the application of stratigraphic contexts of higher resolution and the use of associated radiometric dates related to the most diagnostic archaeological items will allow to improve this approach.

\section*{Acknowledgements}

This paper has been partially supported by grants PID2019-106341GB-I00 and FPU16/00781 from the Ministerio de Ciencia e Innovaci\'on (MCI, Spain),  and grant  AICO/2018/005 from Generalitat Valenciana. JJP is supported by grant FPU16/00781 from the Ministerio de Ciencia e Innovaci\'on and  SPG by Generalitat Valenciana postdoctoral grant APOST-2019/179.

\bibliographystyle{chicago}

\end{document}